\documentclass[journal,comsoc]{IEEEtran}
\usepackage{amsmath,amsfonts}
\usepackage{algorithmic}
\usepackage{algorithm}
\usepackage{array}
\usepackage[caption=false,font=normalsize,labelfont=sf,textfont=sf]{subfig}
\usepackage{textcomp}
\usepackage{stfloats}
\usepackage{url}
\usepackage{verbatim}
\usepackage{graphicx}
\usepackage{cite}
\usepackage{xcolor}
\usepackage{booktabs}
\usepackage{url}
\usepackage{tikz}
\usetikzlibrary{arrows.meta,positioning,shapes.geometric,fit,backgrounds}
\definecolor{cmWarn}{HTML}{C0392B}
\definecolor{cmSim}{HTML}{2A6F97}
\definecolor{cmReal}{HTML}{2D936C}
\definecolor{cmBridge}{HTML}{E76F51}
\definecolor{preprintyellow}{HTML}{FFF200}

\usepackage[hidelinks]{hyperref}
\def\BibTeX{{\rm B\kern-.05em{\sc i\kern-.025em b}\kern-.08em
    T\kern-.1667em\lower.7ex\hbox{E}\kern-.125emX}}
\usepackage{balance}

\usepackage{fancyhdr}
\fancypagestyle{preprintstyle}{%
  \fancyhf{}%
  \fancyhead[C]{\colorbox{preprintyellow}{\textcolor{black}{\footnotesize\sffamily This is the author's pre-print version. Paper under submission to an IEEE venue.}}}%
}

\begin{document}
 
\title{Semantic-Aware Communications in Industrial IoT:\\
What Can Be Tested Today, and What Is Missing?}
 
\author{Rute C. Sofia,~\IEEEmembership{Senior Member,~IEEE}
\thanks{R. C. Sofia is with fortiss GmbH, Munich, Germany (e-mail: sofia@fortiss.org). 
}}

\pagestyle{preprintstyle}
\maketitle
\thispagestyle{preprintstyle}
 
\begin{abstract}
Classical networks treat every bit as equally important, which wastes time and energy in factories and requires extra traffic engineering for critical traffic. Semantic communications send only what a task needs, but how can they be tested under realistic conditions? This paper examines what can be tested today in 6G industrial IoT, that is, networks of factory machines, robots, and sensors, and what is still missing. It introduces \emph{semantic importance}, a score attached to each message that states how much the message matters, and uses it in a context-driven design to derive eight testing needs. Existing simulators, emulators, testbeds, codecs, and learning frameworks are assessed against these needs, which reveals five gaps. We propose a toolchain of four components built around the ns-3 simulator for simulated and emulated scenarios.
\end{abstract}
 
\begin{IEEEkeywords}
semantic communications, 6G, industrial IoT, simulation, ns-3, Edge-Cloud, federated learning
\end{IEEEkeywords}
\section{Introduction}\label{sec:intro}
\IEEEPARstart{I}n a factory, a robot's \textit{"obstacle ahead"} alert and routine status update compete for the same networking resources. Hence, the critical traffic must be protected and preempted. Industrial networks are also becoming increasingly data-centric, to serve applications that are computationally intensive and time-sensitive. Intent-based and information-centric networking has already shifted the focus from host reachability to what the user wants, and 5G service-based architecture exposes network functions as services. Semantic Communication (SemCom) goes one step further: the meaning and context of the data become design inputs, and only the information relevant to a task is sent.

Testing such mechanisms under realistic conditions is difficult because they require tools that can represent meaning and context. This is particularly relevant in Industrial Internet of Things (IIoT) environments, where semantic traffic must coexist with deterministic control traffic, devices such as robots and vehicles are mobile, and inference and learning are distributed across far Edge devices, Edge nodes, and the Cloud. Another difficulty is that the sender and receiver share a knowledge base (KB) of models. When models are updated or a moving device changes its serving node, the two sides may no longer ``understand'' each other, and few tools can reproduce this.

This paper addresses two questions for the design of semantic-aware IIoT networks: what can be tested today and what is missing? It provides the following contributions: i) a tutorial introduction to \emph{semantic importance}, a per-message, context-dependent score: how it is obtained, what it controls, and what it changes at the PHY, MAC, IP OSI layers; ii) a condensed context-driven design in which importance drives both the per-message serving decision and the per-epoch distribution of models; iii) a first assessment of the support that existing simulators, emulators, testbeds, codecs, and learning frameworks may provide for semantic-aware networking, identifying five gaps; and iv) a network simulator version 3 (ns-3) toolchain that can be applied in three different stages of experimentation and carries the importance notion across the networking layers of the OSI stack.

The remainder of this paper is organized as follows. Section~\ref{sec:related} reviews related work. Section~\ref{sec:basics} introduces SemCom and semantic importance, and Section~\ref{sec:usecases} presents the IIoT-relevant 6G use cases that set the requirements. Section~\ref{sec:design} presents our proposal for a semantic importance and context-driven design, with a worked example, and derives eight evaluation needs. Section~\ref{sec:needs} assesses the tools that can meet these needs and identifies gaps. Section~\ref{sec:toolchain} proposes an ns-3-based toolchain of four components, used in three stages of increasing realism, to close these gaps. 
Section~\ref{sec:conclusion} concludes the study with key findings, open challenges, and future work.

\section{Related Work}\label{sec:related}
Chaccour et al.\ present a tutorial vision of SemCom networks that moves from data-driven to knowledge-driven design, with minimal and generalizable semantic representations, causal reasoning, and new semantic metrics~\cite{chaccour2024less}. They do not address how the importance of a message and the version of the shared models reach the network, which is the focus here. Getu et al. survey the research landscape, metrics, and challenges~\cite{getu2024}, while Zhang et al. unify SemCom with semantic edge computing~\cite{zhang2024}. Shi et al. propose a \textit{Federated Edge Intelligence (FEI)} architecture for semantic-aware networking~\cite{shi2021}, which is the starting point of the design in this paper. The artifacts provided in these works are codecs and datasets at link or model level, such as Deep learning enabled Semantic Communications systems (DeepSC)~\cite{xie2021}.

A reading of the literature carried out in the SemComIIoT project~\cite{d42} shows that codecs are mostly trained centrally, with fixed KBs and simulated links; federated training of codecs has been shown only at small scale (5 to 20 clients, image data, static clients). Learned offloading and resource-allocation decisions are mature, but none of the works read selects the serving Edge node from context together with the availability of semantic models or tracks KB versions when a moving device changes its serving node. Importance-like quantities appear under several names: contextual importance of data in C-V2X platooning~\cite{zhang2025platoon} and the control impact of a packet in IIoT~\cite{khan2026push}. These works define the importance of one decision; none follow it from the application to the lower OSI stack layers, or relates it to the model version that produced it.

In the literature search carried out in 2026 for this paper, no survey of simulators or evaluation tools for semantic-aware networking was found. This work does not propose a new SemCom model; it focuses on how semantic-aware networks can be tested.
\section{SemCom and Semantic Importance in a Nutshell}\label{sec:basics}
Weaver distinguished three levels of the communication problem: the \emph{technical} level (how accurately can symbols be transmitted?), the
\emph{semantic} level (how precisely do they convey the intended meaning?)
and the \emph{effectiveness} level (how effectively does the received meaning affect the receiver's conduct?)~\cite{shannon1949}. Classical systems address the first level pf knowledge. SemCom aims at the other two and, in many Deep Learning (DL) based approaches, relies on an encoder and a decoder that share a KB, which in this paper means the background knowledge and the models both sides rely on. Instead of the exact bits, an encoder sends what matters for the task, e.g., a detected defect instead of
the full camera image.

This paper uses one working abstraction to connect SemCom to the network: the \emph{semantic importance} of a message, a single score, e.g., between 0 and 1, that states how much the message matters for the task of its receiver in the current context. A higher value indicates that losing or delaying the message damages the task more. Importance is not a property of the content alone: obstacle detection is critical when a person is on the path and negligible when the area is closed to persons. Therefore, it combines two aspects: the relevance of the content with respect to the shared KB, as judged by the semantic encoder, and the context, which has four parts~\cite{d42}: application (e.g., task, criticality, role in the control loop), computing and network (e.g., battery, link quality, Edge load), physical (e.g., position, mobility, hazards) and social (e.g., people nearby, interactions, safety role). Context-awareness can be derived from policies, an application model, or learned from the application and user behavior over time. A device attaches the result to the message as a tag, so that the Edge and network can use it without decoding.

For the remainder of this paper, three properties are important. First, importance is time-varying, so context must reach the node that assigns the tag "on time". Second, it is estimated by a model and can therefore be inaccurate; an error changes the deadline, energy budget, and tolerated model age, so its effect must be measured. Third, it is meaningful only relative to the shared KB, so a stale KB degrades prioritization and decoding. Unlike Differentiated Services Code Points (DSCP) Time Sensitive Networking (TSN) classes or 5G Quality of Service (QoS) identifiers, which assign a class to a flow in advance, importance is defined per message.
 
\begin{figure}[!ht]
\centering
\begin{tikzpicture}[
  font=\scriptsize\sffamily, >=Stealth,
  box/.style={draw=black!60, rounded corners=1.5pt, align=center, inner sep=2.5pt, fill=white, minimum height=0.55cm},
  inp/.style={box, fill=cmSim!12, text width=2.35cm},
  pol/.style={box, fill=black!6, text width=1.35cm},
  tagb/.style={box, fill=cmBridge!18, draw=cmBridge, thick, text width=1.35cm},
  use/.style={box, fill=cmReal!12, text width=1.05cm, minimum height=0.48cm},
  arr/.style={->, thick, black!65}]
\node[inp] (k) at (0,0.95) {Content relevance\\ to the shared KB};
\node[inp] (c) at (0,-0.15) {Context: application, computing/network, physical, social};
\node[pol] (p) at (3.05,0.4) {Policy};
\node[tagb] (t) at (4.65,0.4) {Importance tag:\\ score, class};
\node[use] (phy) at (6.45,1.5) {PHY};
\node[use] (mac) at (6.45,0.8) {MAC};
\node[use] (ip)  at (6.45,0.1) {IP};
\draw[arr] (k.east) -- ++(0.2,0) |- (p.west);
\draw[arr] (c.east) -- ++(0.2,0) |- (p.west);

\foreach \n in {phy,mac,ip}{\draw[arr] (t.east) -- ++(0.15,0) |- (\n.west);}
\node[font=\scriptsize\sffamily\itshape, text=black!60] at (0,1.65) {Inputs};
\node[font=\scriptsize\sffamily\itshape, text=black!60] at (3.85,1.2) {Per message};
\node[font=\scriptsize\sffamily\itshape, text=black!60] at (6.45,2.0) {Uses};
\end{tikzpicture}
\caption{Semantic importance as a working abstraction: content relevance and context are combined by a policy into a score and class, attached as a tag that PHY, MAC, IP decisions read.}
\label{fig:importance}
\end{figure}
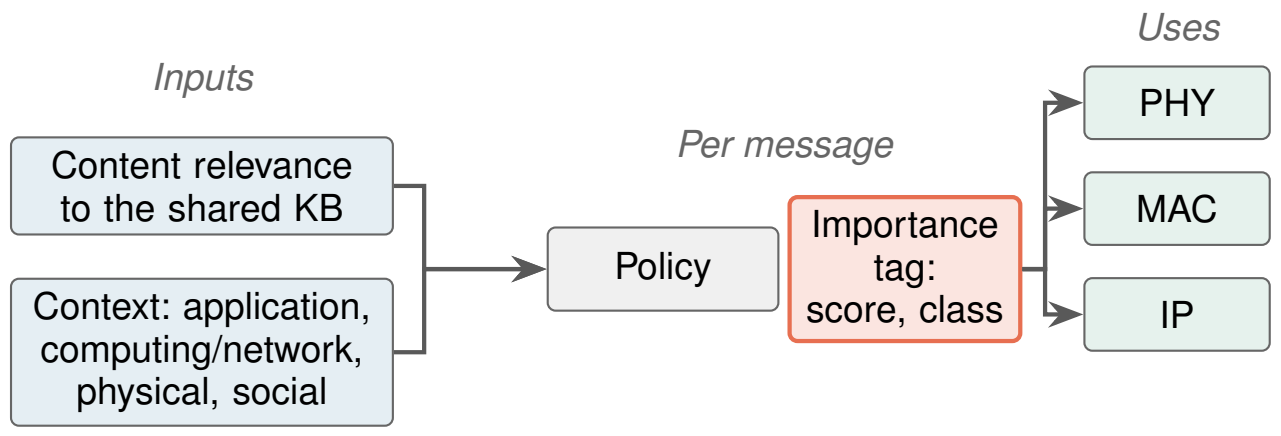

The same tag is used across the OSI stack (Fig.~\ref{fig:importance}). At the \textbf{PHY} layer, the protection level and modulation adapted to convey critical information at low Signal to Noise ratio (SNR), and content-aware error control with graceful degradation. At the \textbf{MAC} layer, priority, scheduling, retransmission, resource allocation by importance, and suppression of redundant transmissions to save energy. In the \textbf{IP} layer, importance can be used as a tag for forwarding, content-based routing, semantic QoS, and drop decisions during congestion.

Hence, every layer of the network stack receives the same importance value from the semantic model, and a mistake in that value affects all of them. For example, a safety alert wrongly marked as unimportant would be sent with weak error protection and low priority, and might arrive too late or not at all. The sender and receiver must also share the same KB, that is, the same version of the models they use to encode and decode messages; otherwise the receiver may misread what was sent. Because the layers depend on each other, testing the impact of semantic importance on each layer is insufficient. A common importance value is only useful if the network can meet the deadlines it implies; therefore, the next section looks at what industrial and 6G use cases require.

\section{6G IIoT Use Cases and Requirements}\label{sec:usecases}

Compared with 5G, 6G targets data rates beyond 1~Tb/s, latency of 10-100~$\mu$s, reliability of up to 99.99999\%, connectivity density of $10^7$ devices/km$^2$ and very high security and privacy. These are vision targets for the most demanding services, and industrial control has its requirements. Among the 6G use-case categories, SemCom is expected to help where data volumes are large, highly correlated, or only partially relevant to the task~\cite{getu2024}.
Table~\ref{tab:usecases} lists 6G use-case categories and their corresponding requirements. Four of these categories are particularly relevant for industrial settings, the focus of this paper:
\begin{itemize}
\item \textbf{Industrial control and robotics:} Small, periodic and time-critical messages for machines and robots, requiring deterministic, bounded latency and low jitter.
\item \textbf{Tactile Internet for remote operations:} From remote
      maintenance to remote surgery, requiring bounded latency and low jitter.
\item \textbf{Digital twins:} Virtual copies that mirror, and sometimes
      control, physical assets. In IIoT, network digital twins are also an important part of the overall IoT-Edge-Cloud digital twinning.
\item \textbf{Cooperative mobile machines (including V2X):} Automated guided vehicles and vehicles that share perception data, where SemCom can help decide when and what to share.
\end{itemize}

The remaining categories, holographic-type communications, multi-sensory
extended reality and space-terrestrial integrated networks, share the need to cut very large data volumes, but are outside the scope of this paper.

In addition to the usual functional networking requirements (bounded latency, low jitter, zero packet loss or low packet loss), in IIoT two additional parameters are relevant. Energy consumption is key, since AI-based mechanisms are energy-hungry, and offloading and local encoding have to be weighed against the energy budget of battery-powered devices. Mobility is also relevant, since moving devices change the serving Edge node and, with it, the models available to a service.
These requirements are, along with the notion of semantic importance, what drives the architectural design debate in the next section.

\begin{table}[ht!]
\centering
\caption{Use-case categories relevant for industrial settings, with indicative
requirements. Figures marked * are 6G vision targets~\cite{d21}.}
\label{tab:usecases}
\footnotesize
\setlength{\tabcolsep}{3pt}
\begin{tabular}{@{}p{0.22\columnwidth}p{0.20\columnwidth}p{0.17\columnwidth}p{0.31\columnwidth}@{}}
\toprule
\textbf{Category} & \textbf{Traffic} & \textbf{Latency} & \textbf{Why SemCom helps}\\
\midrule
Industrial control and robotics & small, periodic & sub-ms to ms & send state changes only\\
\hline
Tactile remote operation & video plus haptics & $<$5 ms* & compress video, protect haptic data\\
\hline
Digital twins & sensor streams, bulk updates & ms* & send only changes that matter\\
\hline
Cooperative mobile machines (V2X) & perception data & 3--100 ms* & decide what and when to share\\
\midrule
Other 6G categories (HTC, XR, STIN) & very high rate & $\mu$s--ms* & outside the scope of this paper\\
\bottomrule
\end{tabular}
\end{table}

\section{Semantic Importance-driven Design}\label{sec:design}
This section outlines a simple design that turns the importance tag into decisions about where a message is encoded, which Edge node serves it and which models each node holds; it is kept deliberately simple so that it can be tested. Three types of entities across the IoT-Edge-Cloud continuum are considered. \emph{Far Edge devices} (sensors, robots, headsets) sense context, attach the importance tag and may encode locally; they are mobile and limited in compute and energy. \emph{Edge nodes} hold a store of versioned semantic models, keep a queue, announce queue and versions to their devices, and encode and decode. The \emph{coordinator} can be placed in the Cloud or at an Edge node, or across multiple Edge nodes. It holds the reference KB, versions the models and distributes model differences.
The importance class of a message sets three limits: its deadline, the energy the device may spend on it (compute, network energy consumption), and its tolerated \emph{staleness}, i.e., by how many versions the model used may be behind the coordinator's. A safety alert has a tight deadline and tolerates no outdated model, whereas routine telemetry has a loose deadline and tolerates a few versions of a model. Decisions are made on two time scales:
\begin{itemize}
\item \textbf{Configuration loop (once per epoch, every few minutes to hours):} determines how learning is organized for a group of devices (centralized, Edge-federated, or fully decentralized) and which model version each Edge node holds. Model differences first go to the nodes whose traffic is the most important, within a backhaul budget per epoch.
\item \textbf{Serving loop (for every message):} lists the options, i.e., each candidate Edge node combined with encoding on the device or at the Edge; estimates time and device energy for each; discards options that miss the deadline, exceed the energy budget, or rely on a model that is too old or missing at the decoding node; and serves the message with the option that uses the smallest combined share of its time and energy limits. If no option remains, the fallback depends on importance: a critical message is sent at once as a model-free message (e.g., a stop command) while the missing model is requested in the background, a low-importance one is served at lower quality, and escalation to the Cloud or a human operator is used only if the deadline allows.
\end{itemize}

The design relies on three assumptions whih can be experimentally verified. \textbf{A1:} urgent messages should not travel to the Cloud, whose round trip may exceed their deadline; the Cloud serves slow background work such as training and KB building. \textbf{A2:} devices should learn locally and share only model updates (federated training) rather than send all data to one place or learn alone; we assume this is the better balance for moving devices and urgent traffic. \textbf{A3:} the best server is not always the nearest: also considering queues, model versions, energy and the movement of the device makes the network miss fewer deadlines, especially for high-importance messages and when the network is busy.
 
Consider an Automated Guided Vehicle (AGV) driving through a factory hall which holds three Edge nodes across the path from Edge~1 towards that of Edge~2, where a second AGV waits. The AGV first \emph{tags} the importance and assesses the path possibilities, namely, three Edge nodes, two places of encoding. The AGV then \emph{estimates} for each the time (uplink, queue, encoding, transport, decoding and the expected cost of a handover) and the AGV's expected energy consumption. Next, it \emph{filters} out options that do not meet the required latency or are served by a model that is too old, since encoder and decoder work like a shared language and different versions may be misread. Finally, it \emph{chooses} the option with the lowest score, the share of the deadline used plus the share of the energy budget used.

The example shows why testing is difficult: the best choice depends on
queues, movement, model versions and energy at the same time, and today's
tools usually look at only one of them.

\subsection{Derived Needs}
This section defines a set of needs. Each one is a property that an evaluation can measure, and we use them to assess existing tools:
\begin{itemize}
\item \textbf{N1 Meaning preservation:} Do the available codecs keep the information the task needs, over realistic radio channels and at low signal quality?
\item \textbf{N2 Traffic differentiation:} Do the PHY, MAC and IP mechanisms treat messages according to their importance when many users share the network?
\item \textbf{N3 Service continuity under movement:} Do mobile devices keep their connection and still find the models they need at the serving node?
\item \textbf{N4 Placement decisions:} Do inference and training run in the right place, and is the right Edge node chosen to serve each device?
\item \textbf{N5 Learning efficiency:} how much traffic, time and energy do central, Edge-federated and fully decentralised training need to reach a target accuracy on industrial data?
\item \textbf{N6 Delay and jitter bounds:} Can semantic traffic stay within delay and jitter limits, for example, within TSN regions, also over wireless links?
\item \textbf{N7 Energy use:} How much energy do semantic inference, training, and transmission need?
\item \textbf{N8 Version consistency:} How far apart can the models of sender and receiver drift when models are updated and devices move?
\end{itemize}

The needs can be checked through measurements such as task accuracy, end-to-end delay, missed deadlines, handovers, the share of messages served with a known model, rounds and traffic needed for learning, energy per task, and the time models need to become consistent.

\section{From Needs to Toolboxes}\label{sec:needs}

\begin{table*}[!htp]
\centering
\caption{Coverage of the needs by classes of tools ($\bullet$ covered,
$\circ$ partial, -- not covered).}
\label{tab:coverage}
\footnotesize
\setlength{\tabcolsep}{4pt}
\begin{tabular}{@{}lcccccccc@{}}
\toprule
\textbf{Tool class} & \textbf{N1} & \textbf{N2} & \textbf{N3} & \textbf{N4} & \textbf{N5} & \textbf{N6} & \textbf{N7} & \textbf{N8}\\
 & Meaning & Traffic diff. & Movement & Placement & Learning & Delay bounds & Energy & Version cons.\\
\midrule
\multicolumn{9}{@{}l}{\emph{Network simulators and emulators}}\\
ns-3 (simulation)                       & -- & $\bullet$ & $\bullet$ & $\circ$ & -- & $\circ$ & $\circ$ & --\\
OMNeT++ with INET (simulation)          & -- & $\bullet$ & $\bullet$ & $\circ$ & -- & $\circ$ & $\circ$ & --\\
ns-3 (emulation mode)                   & -- & $\bullet$ & $\bullet$ & $\circ$ & $\circ$ & $\circ$ & $\circ$ & $\circ$\\
Mininet, Containernet                   & -- & $\circ$ & -- & $\circ$ & $\circ$ & -- & -- & $\circ$\\
\midrule
\multicolumn{9}{@{}l}{\emph{Radio link and semantic codecs}}\\
Link level (Sionna)                     & $\bullet$ & -- & $\circ$ & -- & -- & -- & -- & --\\
Codecs (DeepSC, GenSC-6G)               & $\circ$ & -- & -- & -- & -- & -- & -- & --\\
\midrule
\multicolumn{9}{@{}l}{\emph{Edge, learning and scenarios}}\\
Edge/fog simulators                     & -- & $\circ$ & $\circ$ & $\bullet$ & -- & -- & $\circ$ & --\\
Emulators and testbeds                  & -- & $\circ$ & -- & $\bullet$ & $\circ$ & $\circ$ & $\bullet$ & $\circ$\\
Federated-learning frameworks           & -- & -- & -- & -- & $\bullet$ & -- & $\circ$ & $\circ$\\
Mobility and robotics (SUMO, ROS 2)     & -- & -- & $\bullet$ & -- & -- & -- & -- & --\\
\bottomrule
\end{tabular}
\end{table*}

Table~\ref{tab:coverage} rates each need
as covered (the tool supports it natively), partial (it needs an abstract model, an extension or extra code) or not covered. A complete list of references, omitted due to the lack of space, is available in~\cite{d42}.

\subsection{Network Simulators and Emulators}
ns-3~\cite{ns3} is a widely used open-source tool that can work as a simulator and as an emulator. In simulation mode, everything is modelled and time is virtual, so runs are fast and repeatable. ns-3 offers add-on modules for 5G (5G-LENA) and Wi-Fi, models for device movement, and models for
the energy used by radios, which suits N2 and N3. Its DetNetWiFi extension adds deterministic Wi-Fi, relevant for N6. ns3-gym and ns3-ai connect it to Python learning tools, so a learned policy can control the simulated network.

In emulation mode, ns-3 connects the simulation to real systems: a real computer, virtual machine or container, for example one running a semantic
encoder or an inference service, behaves as if it were a node of the simulated wireless network. This gives realistic processing times and real software
behaviour. The main limit in both modes is that ns-3 has no notion of semantic meaning: a packet has a size and a priority, not an importance, so importance has to be added as an attribute.
OMNeT++~\cite{omnet} with INET is a comparable simulator. INET includes TSN, the standard for guaranteed delivery times in industry, and Simu5G adds 5G. It has no built-in learning interface like ns-3's. Community projects connect it to Python, such as Veins-Gym and RayNet, but
they target Reinforcement Learning (RL).

Mininet is a lightweight emulator on one computer. Hosts are processes connected by virtual wired links whose speed, delay and loss can be
set, and it is mainly used to test real software and Software-Defined Networking (SDN) scenarios. Any Python or learning code runs unchanged inside it. Containernet extends it so that hosts are Docker containers, which can package a complete semantic encoder, inference service or federated-learning client, so it is a
candidate for the Edge-Cloud testbed. 

Unlike ns-3, both lack radio and movement models (Mininet-WiFi adds wireless stations), and since all hosts share one machine, their timing is not guaranteed. 

\subsection{Radio-link Tools}
A link-level simulator imitates one radio connection in detail. Sionna~\cite{sionna} is an open-source, GPU-accelerated library for 5G and 6G radio research. It is
differentiable, which allows transmitters and receivers to be trained end to
end, and it includes channel models and ray tracing. It is therefore a good
environment to train and test semantic codecs over realistic channels (N1).
It does not model the network or Edge-Cloud behaviour. Proprietary
alternatives exist, such as the MATLAB 5G Toolbox.

\subsection{Semantic Codecs}
Open reference implementations of DL SemCom are available. DeepSC~\cite{xie2021} and its variants for text and speech have public code. GenSC-6G is a testbed that combines generative AI, large AI models and hybrid quantum-classical computing. It releases a dataset and a
modular code base in which encoders, communication modules and task decoders can be exchanged, with case studies on classification, semantic localization and edge language-model tasks. It is the closest open framework to what N1 requires. Both are tested mostly on simple noise and fading channels with a
fixed knowledge base, and neither has a network, movement or Edge-Cloud layer. They complement the network tools but do not replace them. 

\subsection{Edge, Learning and scenario tools.} Edge/fog simulators (CloudSim, iFogSim2, EdgeCloudSim, PureEdgeSim, YAFS) study task placement (N4) but treat a task as a size and a deadline, not as a neural model, with simple network layers and no learning. Emulators and testbeds (Fogify, Containerlab, a Kubernetes far Edge cluster on ARM devices with the CODECO framework~\cite{codeco}, FIT IoT-LAB, OpenAirInterface, srsRAN) measure real latency and energy (N4, N7) but are limited by scale, cost and poor repeatability of radio. Flower~\cite{flower}, FedScale and others measure training rounds, accuracy and communication cost (N5) over an abstract network. SUMO and ROS~2 with Gazebo supply movement and robot workloads (N3); meaningful workloads include defect images, bearing vibration data, and traffic classification data.

\subsection{What is missing}
Each part of the problem is served, at least partially, by a tool. What is missing is the connection between the tools, especially between the networking and the AI sides. We identify five gaps:
\begin{itemize}
\item \textbf{G1, Meaning and network are separate.} Codecs are tested on simple channels, while network tools see packets by type, size and priority, not by importance or quality.
\item \textbf{G2, Learning, movement and placement are separate.} Learning tools ignore the network, and Edge simulators ignore learning, so the effect of model updates on traffic, mobility and placement is not studied.
\item \textbf{G3, Model versions are not tracked.} No tool models directly which model version sender and receiver hold, yet this decides which node may serve a message. Staleness can only be measured with added instrumentation.
\item \textbf{G4, Small devices, timing and energy are usually modelled.} Emulators give no timing guarantees, containers only approximate small far Edge devices, and no tool models the energy used by semantic encoders and decoders.
\item \textbf{G5, There is no shared benchmark.} Scenarios, workloads and metrics differ between studies, so results cannot be compared.
\end{itemize}

In short, no single tool meets all the needs. ns-3 is the most flexible base for the network part, because it combines wireless, mobility and energy models with interfaces to AI/ML tooling, and an emulation mode in which real software takes part. It still needs a codec layer  and a learning framework, and it must be extended with importance, model versions and energy consumption at the node and network level. Section~\ref{sec:toolchain} shows how.
 
\section{Closing the Gaps: Toolchain}\label{sec:toolchain}
We propose a toolchain based on ns-3, which already provides Wi-Fi, mobility, queue, and energy models and interfaces to Python, with four components:
\begin{itemize} 
\item \textbf{C1, semantic codec (G1).} Codecs are trained in PyTorch over realistic channels from Sionna. The result is a set of quality tables: how often a task is still solved correctly at each signal quality, message size and model-version difference. ns-3 uses them as a lookup, since for each received message it knows the signal quality and size.
\item \textbf{C2, network (basis).} ns-3 carries every message and maintains the clock. It carries importance as a packet tag (a new tag class), mapped to the access category (MAC) and to a DSCP class with a priority queue (IP), and models many users, movement, radio energy and deterministic Wi-Fi with extensions.
\item \textbf{C3, learning (G2).} Flower or FedScale train encoders within three degrees of decentralization. Model-update traffic and delay are integrated into ns-3, so learning competes with ordinary traffic and the resulting delays feed back to the learning side.
\item \textbf{C4, Edge-Cloud testbed (G3, G4).} A small Kubernetes cluster of far Edge devices and a containerlab-based software copy of the network, with CODECO observability, measure delay and energy, which are returned to ns-3 as parameters. Model versions are tracked in a version table per Edge node and signal type: an ns-3 extension in the first two stages, instrumented in the containers in the third stage.
\end{itemize}

\begin{figure*}[!ht]
\centering
\begin{tikzpicture}[scale=0.72, transform shape,
  font=\scriptsize\sffamily, >={Stealth[length=1.5mm]}, line cap=round,
  b/.style={draw=cmSim, fill=cmSim!10, rounded corners=2pt, align=center, inner sep=2pt, minimum height=0.62cm, text width=3.6cm},
  tg/.style={b, draw=cmBridge, fill=cmBridge!18, thick},
  net/.style={b, draw=black!55, fill=black!6},
  ed/.style={b, draw=cmReal, fill=cmReal!12},
  ext/.style={draw=black!45, densely dashed, fill=black!5, rounded corners=2pt, align=center, inner sep=2pt, minimum height=0.55cm},
  lab/.style={font=\tiny\sffamily, text=black!70, inner sep=1pt, fill=white, fill opacity=0.85, text opacity=1},
  hd/.style={font=\scriptsize\bfseries\sffamily},
  fl/.style={->, draw=black!60, line width=0.6pt}]
\begin{scope}[on background layer]
  \fill[cmSim!6,  rounded corners=5pt] (-2.2,-2.35) rectangle (2.2,4.1);
  \fill[black!4,  rounded corners=5pt] (2.6,-2.35) rectangle (7.0,4.1);
  \fill[cmReal!7, rounded corners=5pt] (7.4,-2.35) rectangle (11.8,4.1);
  \fill[black!3, rounded corners=5pt] (-2.2,-3.55) rectangle (11.8,-2.75);
\end{scope}
\node[hd, text=cmSim!80!black]  at (0,3.85) {Device (AGV)};
\node[hd, text=black!70] at (4.8,3.85) {Wi-Fi, IP, backhaul};
\node[hd, text=cmReal!70!black] at (9.6,3.85) {Edge nodes, coordinator};
\node[b]  (cp)  at (0,3.05) {\textbf{ContextProvider}\\trace sinks: mobility, PHY, queue, energy};
\node[b]  (ie)  at (0,1.95) {\textbf{ImportanceEstimator}\\policy object, score and class};
\node[tg] (st)  at (0,0.85) {\textbf{SemanticTag} (\texttt{Tag})\\+ \textbf{SemanticHeader} (\texttt{Header})};
\node[b]  (sel) at (0,-0.35) {\textbf{ServingSelector}\\in \textbf{SemanticSenderApp} (\texttt{Application})};
\node[b]  (en)  at (0,-1.55) {\textbf{CodecEnergyModel} (\texttt{DeviceEnergyModel})\\+ \texttt{WifiRadioEnergyModel}};
\node[net] (cm) at (4.8,3.05) {\textbf{ImportanceClassMapper}\\\texttt{SocketIpTosTag}, \texttt{SocketPriorityTag}};
\node[net] (pf) at (4.8,1.95) {\textbf{ImportancePacketFilter}\\(\texttt{PacketFilter}) + \texttt{PrioQueueDisc}};
\node[net] (wf) at (4.8,0.85) {Wi-Fi: DSCP to user priority to\\EDCA access category (\texttt{WifiMac})};
\node[net] (bh) at (4.8,-0.35) {Backhaul: \texttt{PointToPointNetDevice}};
\node[net] (mo) at (4.8,-1.55) {\texttt{MobilityModel}, \texttt{StaWifiMac}\\traces \texttt{Assoc}, \texttt{DeAssoc}};
\node[ed] (es) at (9.6,3.05) {\textbf{EdgeServiceApp}\\\textbf{ModelStore}, queue, encode/decode};
\node[ed] (sa) at (9.6,1.95) {\textbf{StateAnnouncer}\\periodic UDP: queue, versions};
\node[ed] (co) at (9.6,0.85) {\textbf{CoordinatorApp}\\epoch loop (\texttt{Simulator::Schedule})};
\node[ed] (df) at (9.6,-0.35) {Model differences:\\\texttt{BulkSendApplication}, most important first};
\node[ed] (ct) at (9.6,-1.55) {\textbf{CodecTable}\\lookup: SNR, size, version difference};
\node[font=\scriptsize\sffamily] at (4.8,-3.15) {\textbf{SemanticStats}: reads the tag at send and receive, \texttt{QueueDisc} and \texttt{StaWifiMac} traces, energy source; writes one record per message};
\draw[fl] (cp) -- (ie);
\draw[fl] (ie) -- (st);
\draw[fl] (st) -- (sel);
\draw[fl] (sel) -- (en);
\draw[fl] (sel.east) -- ++(0.55,0) |- (cm.west);
\draw[fl] (cm) -- (pf);
\draw[fl] (pf) -- (wf);
\draw[fl] (wf.east) -- (7.3,0.85) |- (es.west);
\draw[fl] (sa.west) -- (7.1,1.95) -- (7.1,-0.95) -- (1.0,-0.95) -- (1.0,-0.66);
\node[lab] at (3.2,-0.95) {announced state};
\draw[fl] (co) -- (sa);
\draw[fl] (co) -- (df);
\draw[fl] (df.west) -- (bh.east);
\draw[fl] (ct.north) -- ++(0,0.0) -- (df.south);
\draw[fl, densely dashed] (mo.north) -- (bh.south);
\node[ext] (su) at (-1.2,5.0) {SUMO to \texttt{Ns2MobilityHelper}};
\node[ext] (c1) at (3.4,5.0) {C1 codec tables (CSV)/YAML config};
\node[ext] (ai) at (8.4,5.0) {C3 Python agent, Flower\\via \texttt{OpenGymEnv}};
\draw[fl] (su.south) -- node[lab, right] {I1} (-1.2,4.1);
\draw[fl, densely dashed] (c1.south) -- node[lab, right] {I2} (3.4,4.1);
\draw[<->, draw=black!60, line width=0.6pt, densely dashed] (ai.south) -- node[lab, right] {I3, I4} (8.4,4.1);
\node[ext, draw=cmReal, fill=cmReal!10] (tb) at (13.1,0.85) {C4 testbed\\\texttt{TapBridge}\\proxy nodes};
\draw[<->, draw=cmBridge, line width=1.2pt] (tb.west) -- node[lab, above] {I5, I6} (11.8,0.85);
\end{tikzpicture}
\caption{Proposed \texttt{semcom} module and the ns-3 classes it extends or uses. Orange: the importance tag, created by the device and read by every later block; blue: device; gray: existing network stack with new classifier and mapper; green: Edge nodes and coordinator. Dashed boxes represent inputs: mobility traces (I1), codec tables (I2), a Python agent through \texttt{OpenGymEnv} (I3), a learning framework (I4); the testbed attaches through bridges (I5, I6).}
\label{fig:ns3design}
\end{figure*}
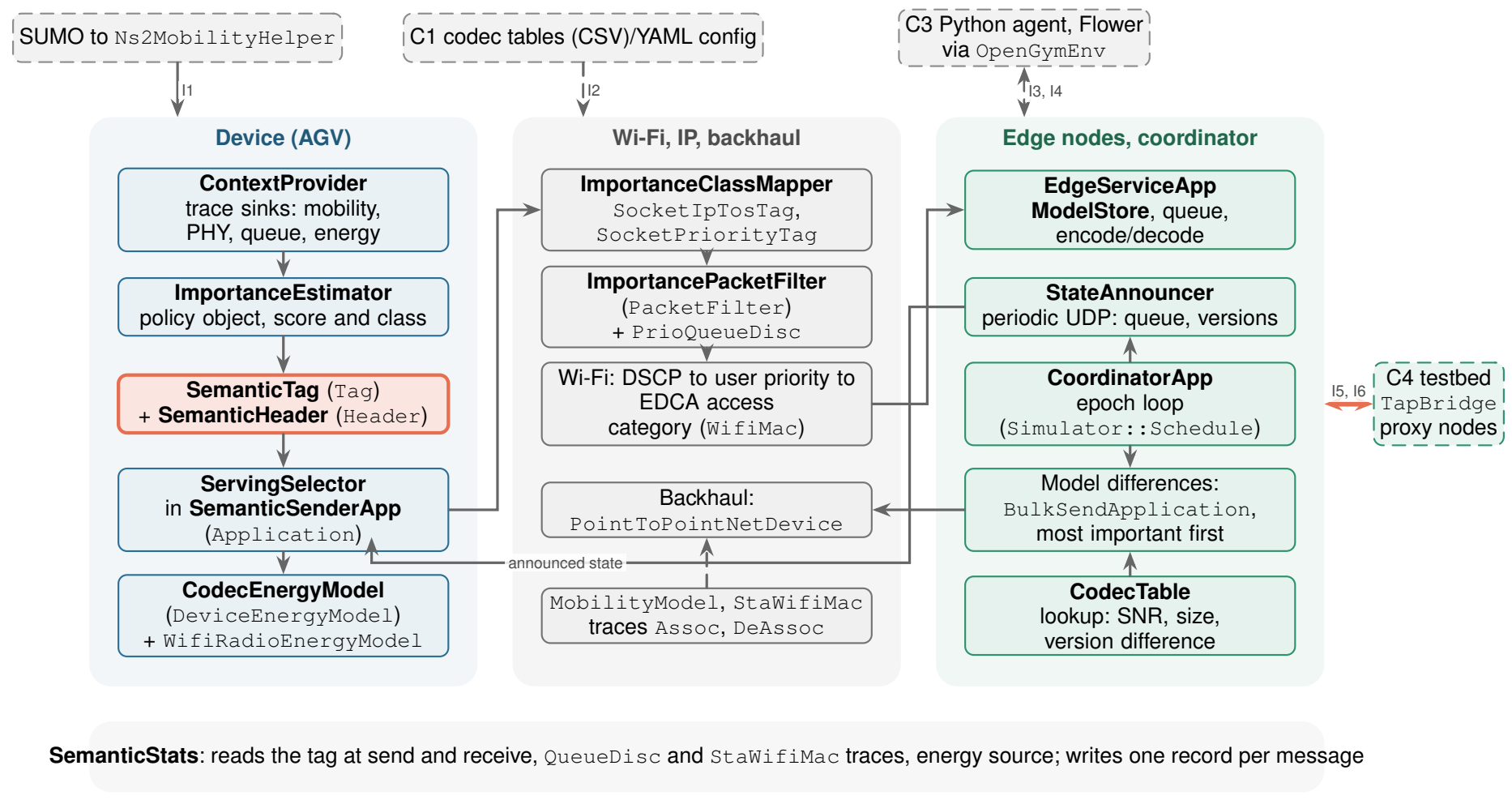
 
\textbf{Importance block.} On each far Edge device (the AGV in our example), a \textit{ContextProvider} reads the four context parts from existing trace sources (mobility model, Wi-Fi PHY, queue disc, energy source) and from the scenario script. An \textit{ImportanceEstimator} applies an interchangeable policy (rule-based, table-based, or learned in Python) and returns the score and class. A \textit{SemanticTag}, a subclass of the ns-3 \texttt{Tag} class, stores both with task, model version, deadline, energy budget, and staleness limit. In Stage 3, a small \textit{SemanticHeader} carries the same fields across the bridge, where the proxy node turns it back into a tag.

\textbf{Use of the tag.} The tag is then read in four places. On the device, the \textit{ServingSelector} estimates delay from the announced queues, the link rate, the handover cost, and the energy use, and picks the serving Edge node. In the network, the \textit{ImportanceClassMapper} turns the class into a DSCP value and a socket priority; the Wi-Fi device maps the priority to an access category, and a classifier with a priority queue disc serves and drops packets by class. At the Edge, each node announces its queue and model versions and decodes with the version named in the tag, and the coordinator updates the models once per epoch and sends the differences over the backhaul. Finally, \textit{SemanticStats} writes one record per message and reports, per class, missed deadlines, high-importance messages served with a stale model, and wrongly tagged messages; per epoch, it reports update traffic, time to consistency, and handovers.

The toolchain can be applied in three stages on the same ns-3 scenario. \textbf{Stage~1, trace-driven:} codec tables and inference, update, and energy parameters measured on the testbed are read by ns-3, every node is modelled, and semantic quality is computed offline; runs are repeatable and faster than real time. \textbf{Stage~2, online-coupled:} quality, model selection, or round control come from a Python process through ns3-ai or ns3-gym, allowing adaptive policies such as learned Edge selection. \textbf{Stage~3, emulated:} the coordinator and, where needed, the Edge nodes and the codec run as real software on the testbed, attached to the simulated network through proxy bridges, so that real processing and update traffic cross the simulated radio and backhaul. A positioning system reports each real AGV's coordinates to its proxy node, so handover and signal quality follow the real trajectory, while the channel remains a calibrated model.
\begin{figure*}[ht!]
\centering
\begin{tikzpicture}[
  font=\scriptsize\sffamily, >={Stealth[length=1.5mm]}, line cap=round,
  x=1.78cm, y=-0.255cm,
  actor/.style={rounded corners=4pt, minimum width=1.55cm, minimum height=0.65cm,
                inner sep=1.5pt, align=center, text=white, draw=none},
  real/.style={actor, fill=cmReal!85},
  sim/.style={actor, fill=cmSim!85},
  oth/.style={actor, fill=black!55},
  life/.style={draw=black!25, densely dashed, line width=0.4pt},
  msg/.style={draw=black!80, line width=0.9pt, ->},
  rsp/.style={draw=black!55, line width=0.8pt, densely dashed, ->},
  warn/.style={draw=cmWarn, line width=0.9pt, ->},
  lab/.style={font=\tiny\sffamily, inner sep=1pt, fill=white, fill opacity=0.9, text opacity=1},
  nt/.style={anchor=west, font=\tiny\sffamily, inner sep=1.5pt, fill=white, fill opacity=0.9, text opacity=1, draw=black!30, rounded corners=1pt},
  stp/.style={circle, fill=black!75, text=white, font=\tiny\bfseries\sffamily,
               minimum size=0.3cm, inner sep=0pt}]
\node[real] (p)  at (0,0) {\textbf{AGV 1}\\[-2pt]estimator};
\node[real] (s)  at (1,0) {\textbf{AGV 1}\\[-2pt]selector};
\node[sim]  (w)  at (2,0) {\textbf{ns-3}\\[-2pt]Wi-Fi MAC IP};
\node[real] (ea) at (3,0) {\textbf{Edge A}\\[-2pt]serving};
\node[sim]  (bh) at (4,0) {\textbf{ns-3}\\[-2pt]backhaul};
\node[real] (eb) at (5,0) {\textbf{Edge B}\\[-2pt]decoder};
\node[real] (co) at (6,0) {\textbf{Coordinator}\\[-2pt]Cloud, FL server};
\node[real] (a2) at (7,0) {\textbf{AGV 2}\\[-2pt]target};
\node[oth]  (ss) at (8,0) {\textbf{Stats}\\[-2pt]SemanticStats};
\foreach \n in {p,s,w,ea,bh,eb,co,a2,ss} \draw[life] (\n.south) -- ++(0,23.2);
\draw[rsp] (6,2.2) -- node[lab, above]{model differences (epoch)} (3,2.2);
\draw[rsp] (3,2.2) -- node[lab, above]{announced state} (1,2.2);
\node[stp] at (-0.32,4.2) {1};
\node[nt] at (0,4.2) {context, score, class, SemanticTag};
\draw[msg] (0,5.4) -- node[lab, above]{tag} (1,5.4);
\node[stp] at (0.68,6.6) {2};
\node[nt] at (1,6.6) {list, estimate, filter, choose};
\node[stp] at (0.68,7.8) {3};
\node[nt] at (1,7.8) {encode on AGV or send raw; header};
\node[stp] at (0.68,9.2) {4};
\draw[msg] (1,9.2) -- node[lab, above]{msg + header} (2,9.2);
\node[nt] at (2,10.3) {DSCP, priority, queue disc, access category};
\draw[rsp] (1,11.4) -- node[lab, above, pos=0.8]{log send} (8,11.4);
\draw[msg] (2,12.6) -- node[lab, above]{radio} (3,12.6);
\node[stp] at (2.68,13.7) {5};
\node[nt] at (3,13.7) {version, queue, encode};
\draw[msg] (3,14.9) -- node[lab, above]{backhaul} (5,14.9);
\node[stp] at (4.68,16.0) {6};
\node[nt] at (5,16.0) {decode with version in tag};
\draw[rsp] (5,17.3) -- node[lab, above]{decoded result} (7,17.3);
\draw[warn] (5,18.8) -- node[lab, above, text=cmWarn]{no version: stop} (7,18.8);
\draw[warn] (5,20.0) -- node[lab, above, text=cmWarn]{get model} (6,20.0);
\node[stp] at (6.68,21.3) {7};
\node[nt] at (7,21.3) {react};
\draw[rsp] (7,22.5) -- node[lab, above]{log receive} (8,22.5);
\end{tikzpicture}
\caption{Communication sequence for the AGV example. Green: real testbed components; blue: simulated by ns-3; gray: metrics collector. Solid arrows: messages; dashed: state, results, and logs; red: fallback if the decoder lacks the version named in the tag.}
\label{fig:sequence}
\end{figure*}
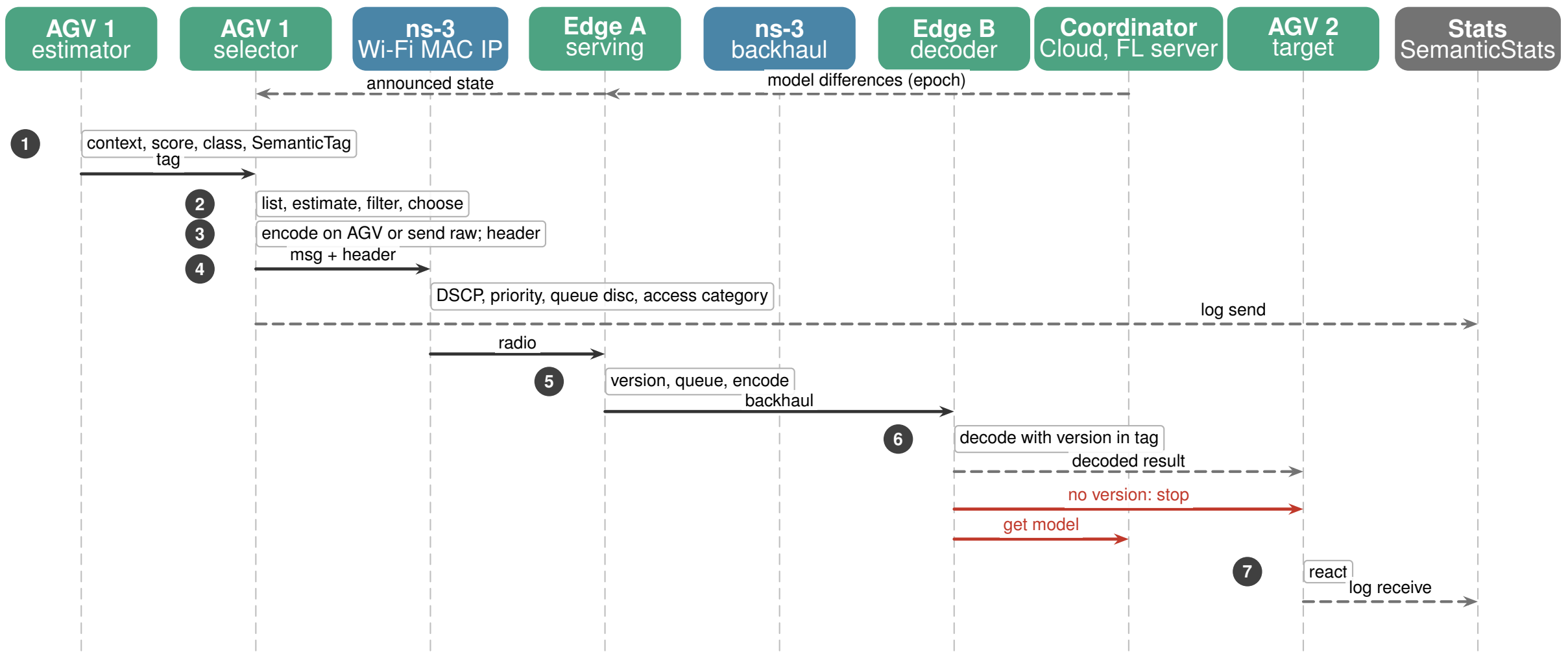

As an example, consider the case of AGVs and obstacle alerts. AGV~1 (source) and AGV~2 (destination) run the Robot Operating System 2 (ROS~2). Each AGV carries a lightweight encoder copy and writes its version into the tag of every message; Edge nodes hold the full encoder and decoder and train as Flower clients, and the Cloud aggregates and receives escalations. Figure~\ref{fig:sequence} then shows the steps of an alert:
\begin{enumerate}
\item On AGV~1, the ContextProvider and ImportanceEstimator detect a person in the path; the policy and the context (speed, distance, people present) give the highest class, stored in a SemanticTag.
\item The ServingSelector lists the options, estimates time and energy from the announced state, discards the infeasible ones, and picks the lowest score.
\item The encoder keeps what matters (``obstacle, type, position, distance'') instead of the whole frame, or the raw frame is sent for encoding at the Edge; the importance fields go into the SemanticHeader.
\item The message enters ns-3 through the bridge, where the tag is restored and logged; the ImportanceClassMapper sets DSCP and priority, so urgent messages are served first on the simulated radio.
\item The serving Edge node checks the required version, queues the message, runs inference, and forwards it over the simulated backhaul.
\item The Edge node of AGV~2 (or AGV-2) decodes with the version named in the tag.
\item A ROS~2 node on AGV~2 publishes the obstacle and its navigation reacts; the result is logged.
\end{enumerate}

\section{Conclusions and Future Work}\label{sec:conclusion}
This paper discussed what can be tested today in semantic-aware networking for 6G industrial IoT. It defined semantic importance as a per-message, context-dependent score that sets the deadline, energy budget and tolerated model age of a message and is read by the PHY, MAC, IP and Edge levels, and a simple design with two decision loops and three assumptions that experiments can check. Eight needs were derived from industrial use cases. Each piece of the design can be tested on its own, but no tool links the quality of meaning to network behaviour, treats learning, movement and placement together, tracks which model version sender and receiver hold, captures the timing and energy of AI on small devices, or offers a common benchmark. Therefore, we propose a toolchain of four components around ns-3, used in three stages: trace-driven with measured tables, online-coupled with Python agents, and emulated with real vehicles, Edge nodes and coordinator attached to the simulated network.
 
However,several challenges remain to be addressed. Keeping the sender and receiver models consistent is not supported by any tool, nor are protocols for exchanging models securely. Importance can be incorrect, because it is estimated by a model, and can be manipulated, since a compromised device could mark every message as critical; how well the serving loop tolerates importance errors, and how tags can be authenticated, are untested. The fallback rules for messages that cannot be decoded must be tested. Fully decentralized learning, without a coordinator, is not supported by the frameworks considered. Guaranteed bounded delay can be simulated but only verified on real hardware, and real-time emulation limits the size of the scenarios.
 
Future work will proceed in four steps: building the connections between the tools (importance attribute and its mapping to access classes, model versioning and a computation-energy extension in ns-3); measuring on an existing testbed the values that the simulation needs, such as inference time, energy per inference, update size and the quality loss of older models; verifying the tool ratings by hands-on testing and confirming tool details such as the deterministic Wi-Fi extension, versions and licences; and developing validation scenarios, workloads and metrics that can serve as a first common benchmark for semantic-aware IIoT.

\section*{Acknowledgment}
This work was funded by the NSFC-DFG Mobility Programme ``Semantic Communications for IIoT'' (M-0626). 

\bibliographystyle{IEEEtran}
\bibliography{references}

\begin{IEEEbiography}[{\includegraphics[width=1in,height=1.25in,clip,keepaspectratio]{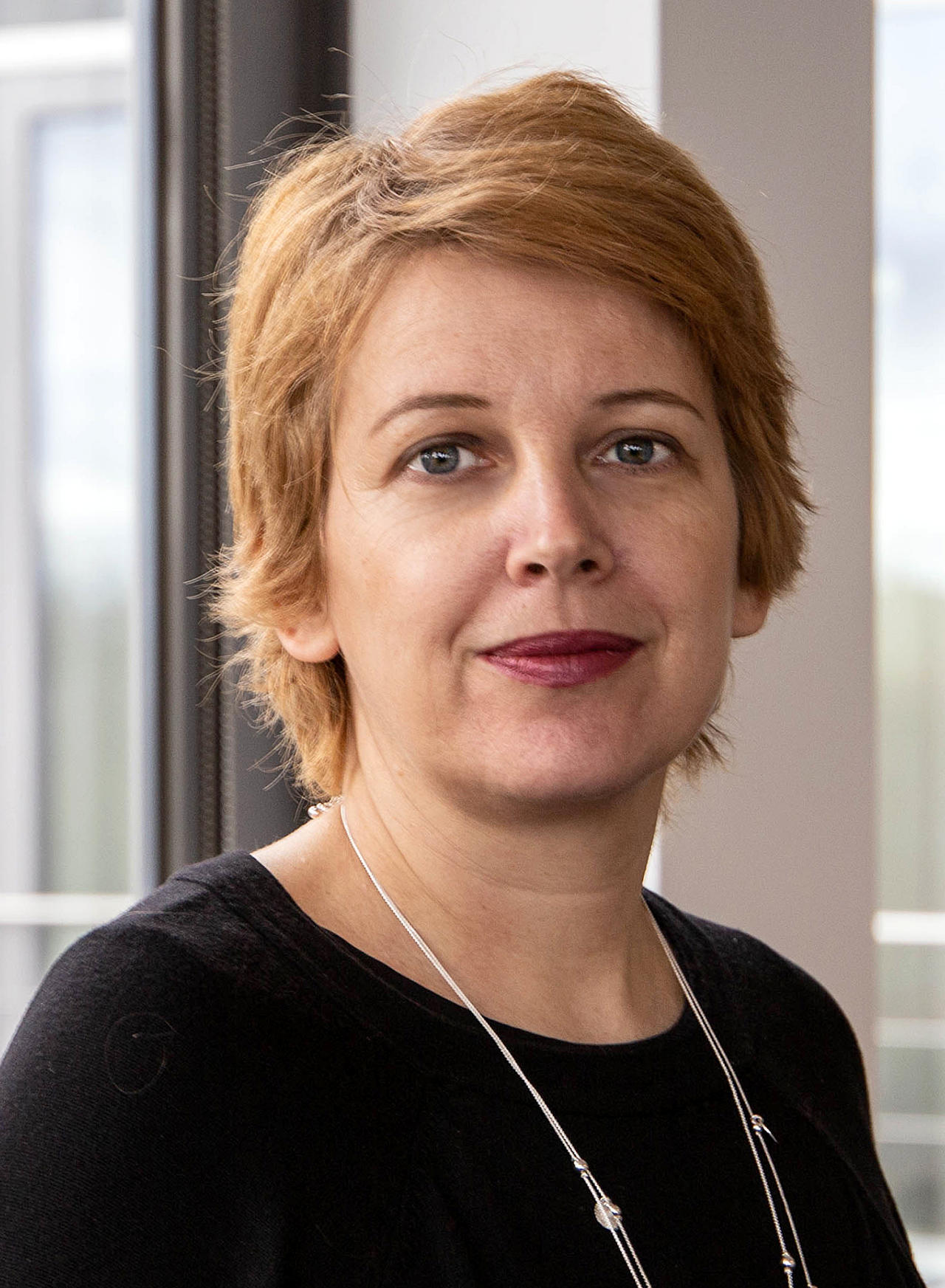}}]{\textsc{Rute C. Sofia}} (Ph.D. 04) heads the Industrial IoT competence field at fortiss, Munich, Germany, and is an invited Associate Professor at Universidade Lus\'ofona. Earlier, she was a Senior Research Scientist at Bundeswehr Universität, Munich; Siemens and Nokia Siemens Networks, Munich; co-led the Internet Architectures and Networking area at INESC TEC, Porto; and co-founded the COPELABS research unit and the startup Senception, Lisbon. She coordinates the 6G CONASENSE platform, co-chairs the energy working group of AIOTI, and serves as a Board Member on the Data Communications Industry (DIA) association. Her research interests include networking architectures and protocols, IoT, Edge computing, AI and networking. She has authored more than 90 peer-reviewed publications and holds eight patents.
\end{IEEEbiography}
\balance
\end{document}